# Electric field enhancement with plasmonic colloidal nanoantennas excited by a silicon nitride waveguide

**Mahsa Darvishzadeh-Varcheie[1], Caner Guclu[1], Regina Ragan[2], Ozdal Boyraz[1] and Filippo Capolino[1,*]**

[1]*Department of Electrical Engineering and Computer Science, University of California, Irvine, CA, 92697, USA*
[2]*Department of Chemical Engineering and Material Science, University of California, Irvine, CA, 92697, USA*
[*]*f.capolino@uci.edu*

**Abstract:** We investigate the feasibility of CMOS-compatible optical structures to develop novel integrated spectroscopy systems. We show that local field enhancement is achievable utilizing dimers of plasmonic nanospheres that can be assembled from colloidal solutions on top of a CMOS-compatible optical waveguide. The resonant dimer nanoantennas are excited by modes guided in the integrated silicon nitride waveguide. Simulations show that 100 fold electric field enhancement builds up in the dimer gap as compared to the waveguide evanescent field amplitude at the same location. We investigate how the field enhancement depends on dimer location, orientation, distance and excited waveguide modes.

## 1. Introduction

The use of nano photonic sensors enable on-site real-time qualitative detection of [1-5] molecules or bio chemical agents with detection limit approaching to single molecule level, and hence leads also to deep understanding of biological and chemical processes that are useful for medical diagnosis [6-11]. Using metal nanostructures for this purpose has been the subject of many studies [12-14] due to the strong near-field plasmonic coupling and local field enhancement that arise between nanospheres in such configurations. Moreover, metal nano architectures may lead to strong surface enhanced Raman scattering (SERS) intensities [15-17] depending on parameters such as size, shape, morphology, arrangement, and local environment. In particular, SERS is an optical spectroscopy method that can use nanoantenna surfaces to enhance signals of light scattered from molecular vibrations for detection of trace amounts of chemical and biological agents [18]. Surface plasmon excitation of a metallic structure in the vicinity of sample molecules is a key factor in the enhanced detection limits in SERS. Surfaces of plasmonic nanosphere clusters have been shown to provide large electric field hot spots [19-22] and they have been utilized in SERS [23-26] where the number of particles per cluster and the cluster separation are stated as important factors for increasing the SERS intensity. Surfaces made of regular arrays made of plasmonic nanospheres have the capability to provide even stronger signals than non-arrayed surfaces [21-22].

Nanosphere-based cluster assemblies in prior studies have exhibited signal enhancements in SERS sensors. However, the challenge of fabricating systems with high reproducibility in SERS response across the sensor surface is not frequently addressed. To overcome this challenge, in our previous work [25],[27] an innovative method for fabricating self-organized clusters of metal nanospheres with nanometer gap spacing on diblock copolymer thin films was successfully investigated. Specifically, monodisperse, colloidal gold nanospheres are attached on chemically functionalized polymer regions on a self-organized diblock copolymer template to engineer a high density of hotspots over large areas while using low cost fabrication methods. Although single molecule detection limits using nanostructured surfaces have been demonstrated previously, issues of large scale fabrication cost, shelf life, reproducibility in detection response, and integration have eluded researchers and this has resulted in limited technological impact.

The goal of this paper is to investigate an on-chip architecture that eliminates structural constraints of an external microscope and integrates the entire excitation and detection mechanism into a single waveguide; thus reducing size and cost, increasing portability, usage flexibility and robustness of future spectroscopy based sensors. On chip spectroscopic measurements are enabled by silicon compatible materials with low loss at visible and at near IR frequencies. Here, the integration of nanoantennas with waveguides that we propose is a spectrometer building block which addresses the above challenges and only requires smaller sample solution volumes in comparison with open systems. With this goal in mind, we investigate the field enhancement capability of plasmonic nanoantennas composed of nanospheres from colloidal solution located on the surface of silicon nitride ($Si_3N_4$) waveguides. Silicon nitride waveguides serve as both illumination and detection channels for exciting and collecting signals from nanoantennas. The silicon nitride waveguide platform is proposed instead of a silicon-on-insulator (SOI) platform for integrated photonic circuits [28-30] due to its advantages such as lower cost fabrication and less propagation loss [31] at both visible and near infrared [32-33]. Moreover in previous work [34-35] the authors fabricated a sub-micron silicon nitride waveguide, experimentally characterized the modal dispersion and loss of the guided wave and demonstrated that silicon nitride can be an alternative to SOI platform. The scheme of assembling dimer nanoantennas on the silicon nitride waveguides from colloidal solution allows for numerous nanoantennas present on the surface of the waveguide with controlled density. We assume in this investigation that dimer nanoantennas on the waveguides are separated by sufficient distances and do not couple with each other such that each nanoantenna can be modeled as an isolated nanoantenna integrated with the waveguide. Hence, in the following we carry out the investigation on the performance of one dimer nanoantenna on top of a waveguide.

The paper is organized as follows. First we demonstrate an *E*-field enhancement results when using the waveguide-driven nanoantenna made of a gold nanosphere dimer and analyze the dependence of the field enhancement on the geometrical parameters. To simulate the performance expected in experimental specimens, we improve the simulated models to include the realistic chemical and physical composition of the fabricated samples. Therefore, another set of models which include a thioctic acid molecular layer (used for assembly,) covering the nanoantenna are adopted next and we investigate how the thickness of this extra layer affects performance. Finally we provide the performance comparison, in terms of field enhancement, between the proposed waveguide-driven nanoantenna structure and the same nanoantenna on top of a multilayer composed of silicon nitride on top of glass, all with the same thicknesses as in the waveguide case, illuminated by a plane wave.

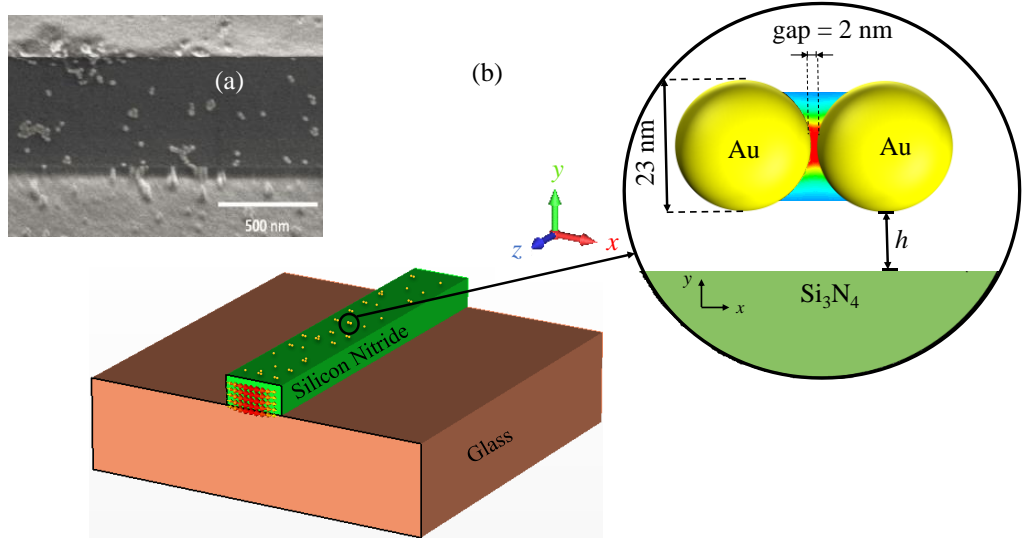

Fig. 1. (a) Scanning electron microscope (SEM) image with gold nanospheres on top of silicon nitride waveguide [36], (b) Proposed integrated CMOS-compatible waveguide with plasmonic nanoantennas (monomers, dimers, etc) on its top surface for field enhancement. In this paper we investigate the optical excitation of a dimer via guided modes in a silicon nitride waveguide, though other waveguide material choices could be utilized as well. On the left, the cross section field distribution of the fundamental propagating waveguide mode is shown, whereas on the right the inset shows the dimensions of the dimer and the "hot spot" where the strongest field occurs as a result of waveguide excitation.

## 2. Field enhancement by a nanoantenna integrated with a waveguide

In previous work, as in [23-26], nanostructured surfaces made of nanospheres were investigated for detections of low molecular concentrations. The electric field enhancement is a critical parameter in decreasing detection limit down to single molecule level. Here we propose the use of a resonant plasmonic nanoantenna integrated with a CMOS-compatible waveguide to provide strong field enhancement. Specifically, the nanoantenna is composed of a dimer of gold nanospheres sitting above a silicon nitride waveguide that also provides the optical excitation as shown in Fig. 1. In this figure, plasmonic nanoantennas (monomers, dimers, trimers, etc) are depicted on top of silicon nitride waveguide. In this article we investigate the field enhancement produced by a dimer on top of the waveguide. In fact such nanoantennas with subwavelength sizes, in most of the cases, can be treated as isolated scatterers that do not couple significantly based on the assumption that dimer density is low on the waveguide surface. Therefore the significant difference between the proposed structure in Fig. 1 and previous work such as [24] is the integrated excitation method; instead of illuminating the structure with an external beam, as in [24], the gold dimer is excited by the tail field of a guided mode propagating inside a waveguide on a glass substrate. For all cases reported in this paper, the vertical distance from the bottom of the nanospheres to the top surface of the silicon nitride waveguide is denoted by $h$ and the shortest distance between the two nanospheres' surfaces is called the "gap", as shown in Fig. 1. According to the SEM images of a fabricated sample reported in [24-27], an experimentally achievable gap between the nanospheres with diameters of 23 nm is 1-2 nm. Accordingly, in this paper we assume a gap spacing of 2 nm and gold nanosphere diameter of 23 nm and both parameters are kept constant for all the cases reported throughout the paper. The permittivity of gold is taken from the experimental data provided in [37]. Full-wave simulations are performed by a commercial software based on frequency domain finite element method (FEM), implemented in CST Microwave Studio by Computer Simulation Technology AG.

In our model we consider a rectangular silicon nitride waveguide with cross section of $1\mu m \times 0.5\mu m$ as in Fig. 1. The waveguide dimensions are determined based on limits in the conventional optical lithography features where $1\mu m$ is the resolution limit due to optical diffraction limit of direct contact lithography [35], since our aim is not to rely solely on electron beam lithography which could achieve smaller feature size. Because of silicon nitride dielectric density, the given waveguide hosts also higher order propagating modes. With the given dimer orientation (nanospheres aligned along the *x*-axis), a field enhancement is achieved when the dimer is excited by horizontally polarized electric field [38], i.e., polarized along the *x*-direction. Moreover when the dimer is located in the middle of the waveguide top surface, we consider here only the case with the horizontal fields present at the dimer gap, when the guided modes with perfect electric conductor (PEC)-symmetry with respect to the *y-z* symmetry plane are excited. Therefore, we consider the first five propagating modes with PEC symmetry shown in Fig. 2. We define the field enhancement (*FE*) as

$$FE = |\mathbf{E}|/|\mathbf{E_0}| \qquad (1)$$

where $|\mathbf{E}|$ is the electric field magnitude at the center of the gold dimer and $|\mathbf{E_0}|$ is the electric field magnitude at the same location in the absence of the dimer nanoantenna. Fig. 2. shows *FE* in the frequency band from 540 THz to 570 THz when the waveguide is excited with the modes whose *E*-field profile matches the above conditions, taking $h = 25$ nm for the vertical dimer position. The reported results shown in Fig. 2 demonstrate that the dimer resonates at 554 THz and leads to a *FE* level of approximately 25.72. This resonance frequency and *FE* value are effectively insensitive to the mode propagating in the waveguide as observed in Fig. 2.

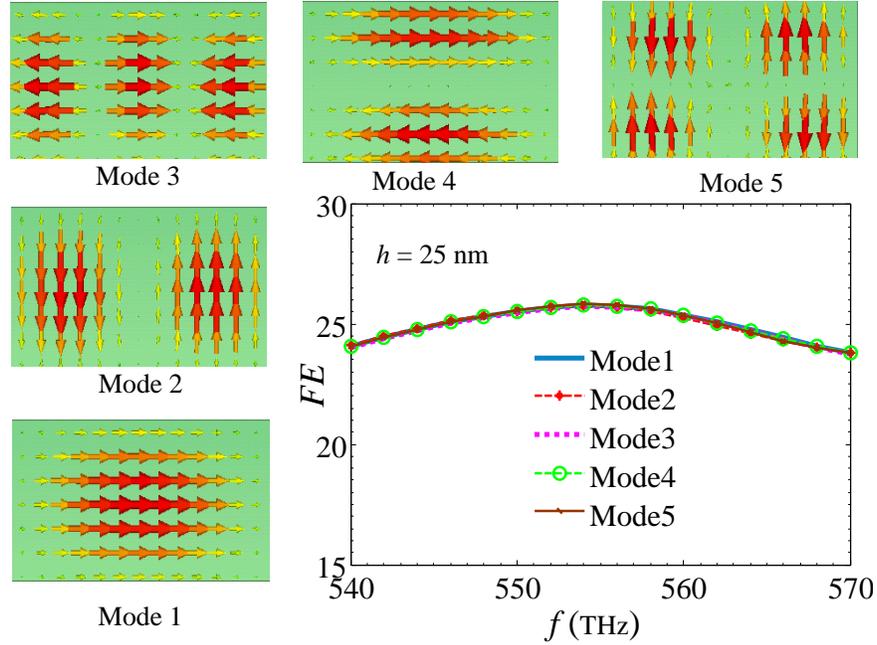

Fig. 2. First five propagating modes (with electric mirror symmetry) in the silicon nitride waveguide, and field enhancement in the middle of the dimer (above the waveguide) when each mode is individually excited.

In the following full-wave simulation results, we assume the waveguide is excited only by its fundamental propagating mode (Mode 1 in Fig. 2.) with horizontal polarization. In order to illustrate the electric field localization, the electric field magnitude profile, normalized to the maximum value occurring at the gap center, is reported on three orthogonal cross-sectional planes of the dimer in Fig. 3. The $E$-field hot spot is localized mainly at the gap as expected, and the field profile between the nanospheres is almost constant indicating the dipolar nature of the scattering from the nanospheres.

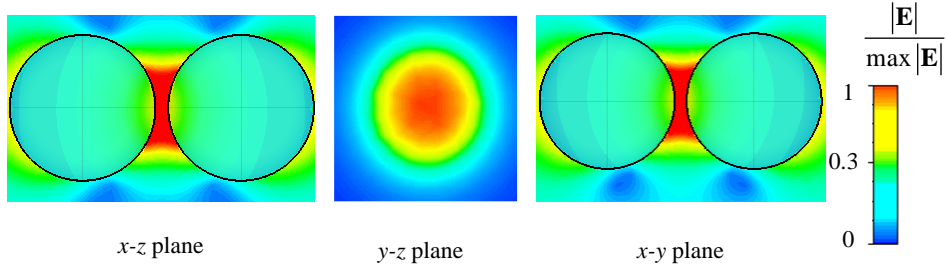

Fig. 3. Electric field magnitude on three principal cross-sectional planes generated by the dimer nanoantenna excited by the fundamental propagating mode of the integrated waveguide, at the dimer's resonance frequency of 554THz. Field has been normalized with respect to its maximum value, and the colors are linearly mapped as shown in the legend. The vertical distance of the dimer from the surface, $h$, is 25 nm.

The $E$-field enhancement profile $|\mathbf{E}(y)|/|\mathbf{E_0}(y)|$ along the $y$ axis (vertical axis in the middle of the waveguide passing through the gap center) is reported in Fig. 4(a), when $h$ is fixed at 25 nm (thus gap center is at $y$ = 286.5 nm) at 554 THz. Here $|\mathbf{E}(y)|$ and $|\mathbf{E_0}(y)|$ denote the magnitude of electric field, both as a function of vertical coordinate $y$ at the presence and absence of the nanoantenna, respectively. (We leave the $y$ dependency in the formula to stress that $|\mathbf{E_0}(y)|$ also depends on vertical position for the analysis presented in Fig. 4(a) whereas in Fig. 4(b) $|\mathbf{E}(y)|$ is normalized with respect to $|\mathbf{E_0}(y=0)|$.) Fig. 4(a) provides the signature of the dimer near field is characterized by a peak at the gap center. The zoom in the inset shows the actual peak width of few nanometers. Despite it is clear that the nanoantenna provides a strong enhancement of the field compared to the field at the same location without a nanontenna. It is also interesting to observe how strong is the field in the nanoantenna gap compared with the strong field guided by the waveguide. Therefore we report this ratio in Fig. 4(b) defined as the magnitude of electric field along a scan line coinciding with the $y$ axis in presence of the nanoantenna, normalized by the electric field magnitude at the center of the waveguide ($y = 0$ nm) where the modal guided electric field is maximum.

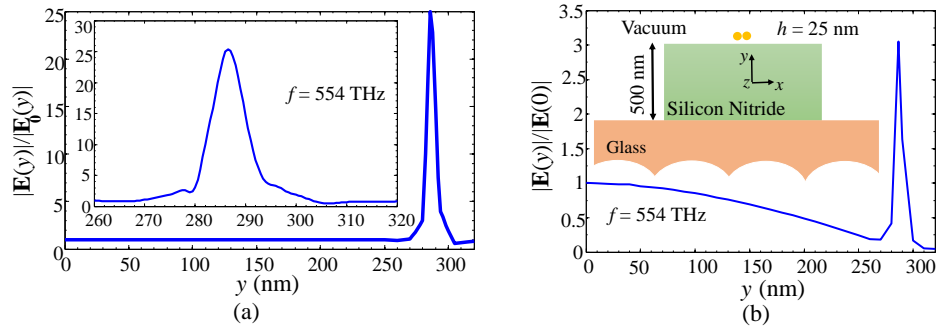

Fig. 4. (a) Field enhancement profile at 554 THz along the $y$ axis, i.e. vertical line passing through the dimer gap center. Here $h = 25$ nm, the dimer is centered at $y = 286.5$ nm where the sharp peak occurs. (b) Electric field profile, normalized by the electric field at the center of the waveguide ($y = 0$ nm) for the same case in (a) at 554 THz. Two things must be noticed: (i) The left plot (a) shows a strong field enhancement within the nanoantenna gap. This field enhancement is further increased if the material hosting the nanoantenna has a larger dielectric constant, as shown next. (ii) The right panel (b) shows that the field within the nanoantenna gap is three times stronger than the electric field at the center of the waveguide.

The electric field at the dimer gap center is three times stronger than the guided electric field at the waveguide center, hereby we stress that the resonant plasmonic dimer's near field, driven by the weak evanescent field outside the waveguide, reaches even larger levels than the maximum of the guided field at the center. Also it should be considered that by invoking reciprocity, the field scattered by the dimer, when excited by an emitting molecule in the gap region, is efficiently coupled into the waveguide and the coupled electric field amplitude is one third of field at the emitted field in the dimer's gap. Note that this result pertains to the well confined field of Mode 1 of the waveguide (Fig. 2.). However one may even speculate that for this kind of applications less confined fields like that of Mode 3 for example could be used to provide even stronger fields in the nanoantenna gap.

Until this point, we simulated the case where a dimer nanoantenna is located exactly above the top center of the waveguide and the gap between them is 2 nm, however in experimental systems, dimers can be located in various regions across the waveguide surface with potentially varying gap spacing. Therefore in the following analysis, we report the field enhancement, $FE$, versus frequency from 400 to 620 THz for various vertical and horizontal positions of the dimer and having different gap spacing in Fig. 5(a), 5(b) and 5(c), respectively. First in Fig. 5(a) the dimer is assumed to be located at various vertical distance from the waveguide surface, $h$, in the range from 0 nm to 25 nm while keeping the horizontal position of the dimer fixed above the center of the waveguide ($x = 0$ nm) and the gap spacing as 2 nm.. As $h$ increases, the field enhancement slightly drops monotonically from 30.4 to 25.7, and the resonance frequency shifts slightly to higher frequencies. In Fig. 5(b) $h$ and the gap are fixed at 25 nm and 2 nm and the dimer's horizontal position is varied with the center of the dimer located at $x = 0$ (the center of the waveguide), 100, 200, 300, 400 nm, keeping in mind that the waveguide terminates at $x = 500$ nm. Results show that the horizontal position of the dimer has negligible impact on the field enhancement level and the resonance frequency except for the position $x = 400$ nm, i.e. where the dimer is closest to the edge of the waveguide. At the edge there is a slight decrease in enhancement level and a small increase in the resonant frequency. The results reported in both Fig. 5(a) and 5(b) indicate that when the average effective dielectric density around the dimer is decreased, a decrease in the enhancement levels and a slight blue shift in resonance frequency is observed. In Fig. 5(c), *FE* versus frequency for different gap size (2, 4, 6 and 10 nm) is reported. According to this figure the highest field enhancement can be obtained in case of a 2 nm gap. Increasing the gap size shifts the resonance frequency slightly (554, 565,567 and 567 THz for the gap size of 2, 4, 6 and 10 nm, respectively). As it can be seen from different parts of Fig. 5., the most critical parameter in our design is the gap size.

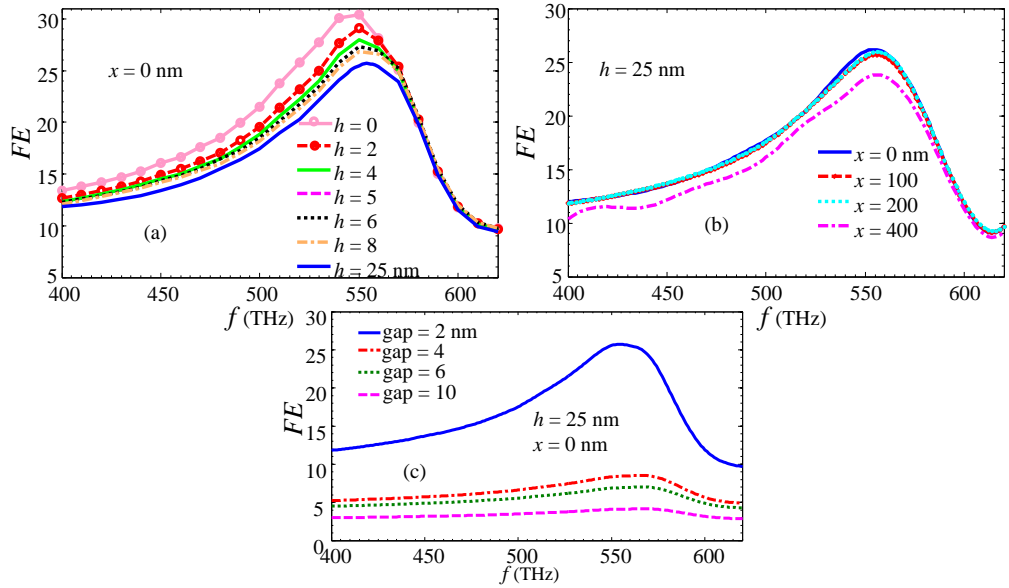

Fig. 5. Field enhancement versus frequency, (a) for the vertical displacements, $h$, when the dimer is above the center of the waveguide and the gap is 2 nm, (b) for various horizontal positions of the dimer when $h = 25$ nm and the gap is 2 nm are kept constant, (c) for various gap size when $h$ is 25 nm and $x = 0$.

Depending on the procedure for depositing dimers on the waveguide surface, one may not have complete control on the orientation of the dimer axis on the *x-z* plane. Thus in Fig. 6., we show the effect of different dimer orientations (90, 60, 30, and 0 degrees with respect to the *z*-axis along the waveguide) on the field enhancements (*FE*) when $h$ is 25 nm. As we expected, *FE* increases when the dimer axis gets closer to the *x*-axis (the axis perpendicular to the waveguide) while the resonance frequency remains constant. Since we have maximum *FE* for the dimer's orientation perpendicular to the $z$ axis ($\theta = 90°$), in all the other simulations we consider only this orientation.

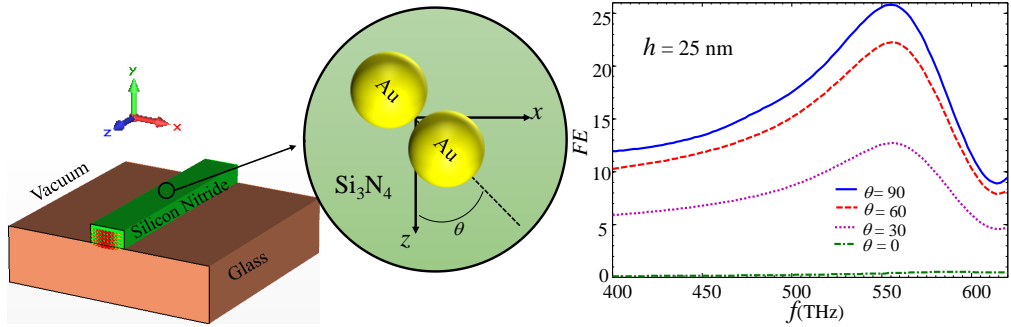

Fig. 6. *FE* for different dimer orientations with respect to the waveguide longitudinal axis *z*.

Here we closely simulate an achievable experimental system. The chemical assembly procedure of [25] and [27] involves depositing gold nanospheres with thioctic acids ligands from colloidal solution on a diblock copolymer thin film composed of poly(methyl methacrylate) and polystyrene (PS-b-PMMA). Nanospheres chemically attach to PMMA domains that are recessed due to the chemical functionalization process [27]. An average dielectric constant of 2.47 results in the region between nanospheres due to PMMA environment and thioctic acid ligands on their surface, creating contrast with respect to the surrounding vacuum and red-shifting the plasmon resonance [24]. These thin film layer on top of the waveguide is referred to here as combination of thioctic acid ligands and PMMA film. Two types of this combination formation models are considered in the following: (Case I) a uniform layer with thickness of 40 nm surrounds a dimer that is partially submerged as shown in Fig. 7 (left panel), and (Case II) a 30nm-thick whole layer together with a 2nm single-molecule layer covering the entire dimer surfaces as in Fig. 7 (right panel). Case I represents the case when the dimers are located in a small valley in the PMMA and then a thin layer of thioctic acid molecules is deposited partially filling the gap region. Case II evaluates the contribution of the thioctic acid monolayer on the nanosphere surface. The comparison of the two cases will provides insight into the enhancement resulting from the PMMA and from the thioctic monolayer. It was reported in [24] that thioctic acid present in the gap has impact on the enhancement and plasmon resonance due to the increased dielectric constant in this region compared to the vacuum. In both Cases I and II we consider the average dielectric constant of the PMMA and the thioctic acid molecules equal to approximately 2.47 for simplicity.

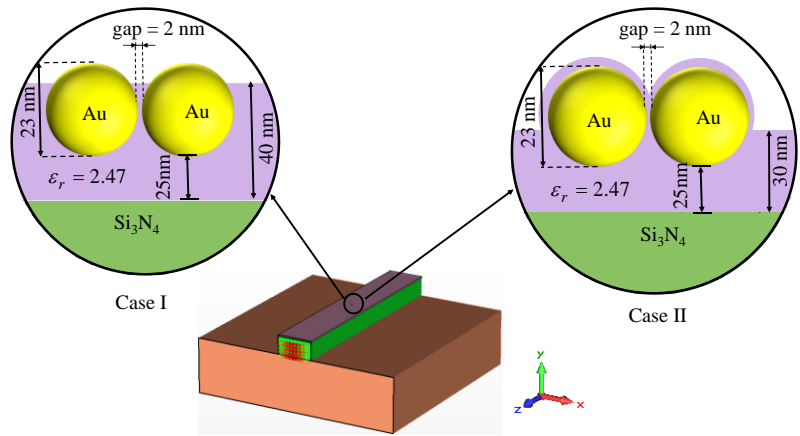

Fig.7. Illustration of the dimer immersed in dielectric layer (represented in purple) covering the waveguide. The two examples, namely Case I and Case II, are investigated.

Fig. 8. shows the comparison of the *E*-field enhancement for Cases I and II illustrated in Fig. 7. and for the case without thin film on the waveguide as depicted in Fig. 1. Note that in Fig. 8(a) $|\mathbf{E_0}|$ in Eq. (1) is calculated when there is no dimer and no thin film (though we have observed that a definition of $|\mathbf{E_0}|$ based on field evaluation considering the thin film, without dimer, would not alter the results). We observe that cases with thin film on waveguide exhibit a larger enhancement peak occurring at a lower frequency. Maximum enhancements are 102, 82, and 25.5 for Case I, Case II and the case without thin film, respectively. On the other hand, at the resonance frequency of the case without the thin film (554 THz), all cases provide almost the same level of enhancement but note this is not the maximum value for the case with the thin film on waveguide. In Fig. 8(b) we show the how the field in the gap region is enhanced with respect to the field at the center of waveguide, for the three cases (I, II, and without the thin film). Our results show that the local field at the gap center is much larger (13.5 for Case I, 10.8 for Case II and 3 for the case without the thin film) than the field at the center of waveguide at resonance frequencies (504, 504 and 554 THz respectively). It is important to understand the dielectric environment can be used as a design parameter for spectroscopy integrated systems in addition to the geometric concepts proposed here.

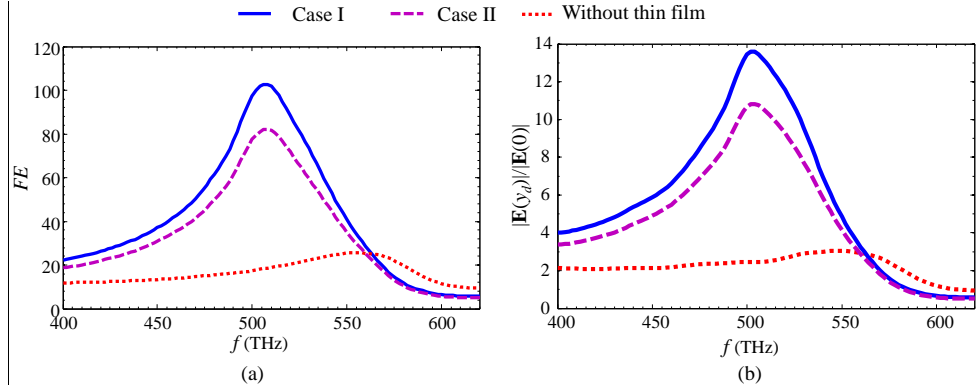

Fig. 8. (a) Comparison of field enhancement and (b) electric field at the dimer center ($y_d$ = 286.5 nm) normalized to field at the waveguide center ($y = 0$), for different environments surrounding the plasmonic dimer nanoantenna: Cases I and II with the thin film on waveguide and the case without it. Here $h$ is 25 nm for all cases.

## 3. Comparison with field enhancement by a nanoantenna excited by a plane wave

In [24], field enhancement resulting from various oligomer geometries of gold nanospheres, including dimers, were investigated on top of layered structure illuminated by an external plane wave. Here, we take the same vertical stack of materials and the dimer nanoantenna as in Section 2 and compare the field enhancement in the layered structure under a plane wave illumination from [24] with the waveguide-driven nanoantenna topology. In Fig. 9. the simulated multilayer structure composed of (from bottom to up) glass, silicon nitride and vacuum is illustrated. The vertical distance between the bottom of the spheres and the top of the silicon nitride layer is denoted by $h$ and the definition of the "gap" follows that in Section 2 as shown in Fig. 9. The silicon nitride layer thickness is $0.5 \mu m$, i.e. equal to the waveguide thickness in Section 2. The two gold nanospheres have constant diameter and gap distance as those in Section 2 (23 nm and 2 nm, respectively). The polarization of normally incident wave is along the *x* direction as shown in Fig. 9. The normal plane wave incidence is simulated assuming a two-dimensional infinitely periodic array along the *x* and *y* dimensions. A single dimer is inside each unit cell which is taken as square in the *x-y* plane with a side length of 375 nm (half a wavelength at 400 THz). In Fig. 9, right panel, we report the field enhancement for various *h* (0nm, 5nm and 25nm) versus frequency ranging from 400 to 620

THz. Our simulation results show that by increasing the distance $h$, the resonance frequency slightly shifts to higher frequencies and field enhancement decreases monotonically, with enhancement levels dropping from 30 to 25. Here it is stressed that similar enhancement levels and resonance frequencies are obtained in Fig. 5. for the waveguide-driven nanoantenna topology shown in Fig. 1. Our simulation results are in agreement with the findings in [24].

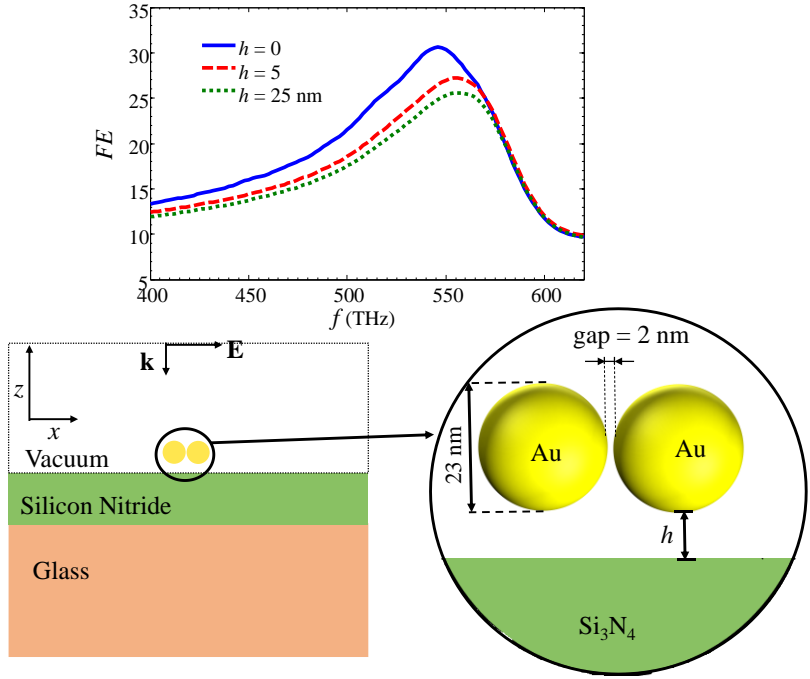

Fig. 9. Topology of a nanoantenna on top of a multilayer structure excited by a normally incident plane wave, and plot of field enhancement versus frequency for $h = 0, 5, 25$ nm, using the same gold nanosphere diameter and gap as in the waveguide-driven structure reported in Section 2.

Next, we assess the impact of the thin film as in Section 2, and we consider the two types of the thin film formation cases as in Fig. 6. but considering plane wave excitation as in Fig. 9. In Fig. 10 we report the results for Cases I and II with the thin film and for the case without the thin film on waveguide for both the waveguide-driven nanoantenna (solid curves) and the plane wave driven multilayer topologies (dashed curves) when $h = 25$ nm. We observe that without the thin film both waveguide-driven and the plane wave driven structures provide the same field enhancement. On the other hand, when considering the cases with the thin film both waveguide-driven structure and the plane wave driven structure provide stronger field enhancement. The waveguide-driven structure exhibits slightly larger field enhancements, $FE$, than the plane wave driven structure (102 and 82 for Case I and Case II versus 95 and 80, respectively). For both the cases of waveguide and plane wave driven structures, when the thin films are taken into account, the resonance frequency of the dimers shift to lower frequencies; yet the frequency shift from the waveguide driven structures is significantly lower than that from the waveguide-driven structures. This may be attributed to mainly the array effects included in the plane wave simulation (approximated by assuming periodic boundary conditions). Overall the waveguide-driven nanoantennas are able to provide similar, even marginally larger, electric field enhancement, $FE$, levels than the plane wave driven structures.

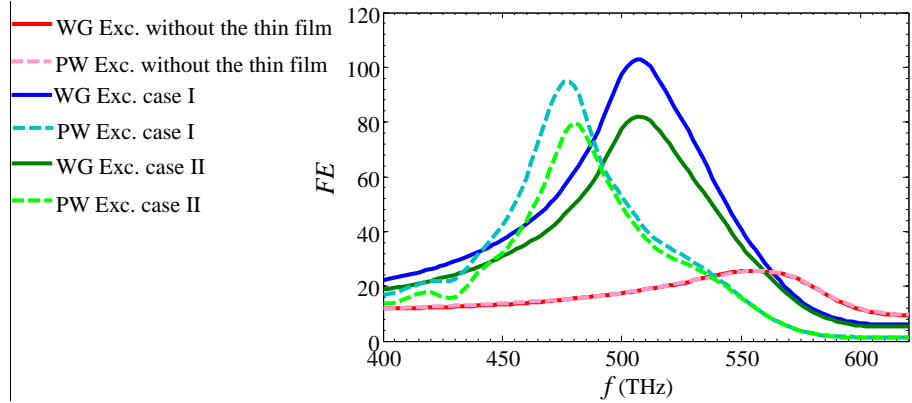

Fig. 10. Comparison of field enhancement for different structures with two kinds of excitation (Exc.). We compare the field enhancement of the waveguide (WG) based topology with the classic one of plane wave (PW) illumination. In both cases the thin film results in stronger field enhancement and in an expected red-shift of the resonance frequency.

## 4. Conclusion

In this paper we have proposed a configuration that provides strong electric field enhancement by placing dimer plasmonic nanoantennas on top of an integrated silicon nitride waveguide. We also considered the effect of a layer of PMMA and thioctic acid molecules which has been used previously for fabrication of dimers assembled from colloidal solution. It is significant that the presence of this polymer film results in an even stronger field enhancement due to an increase in dielectric constant. The results presented here demonstrate that a CMOS compatible waveguide geometry can yield strong field enhancement when inexpensive and chemically driven fabrication processes are adopted. While we consider only the excitation of colloidal nanoantennas in this paper, the results imply that scattering process by molecules in the nanoantenna gap would also generate the excitation of guided modes that can be eventually detected in silicon technology.

**Acknowledgments**

We acknowledge partial support from the National Science Foundation, NSF-SNM-1449397.
    We are grateful to CST Simulation Technology AG for letting us use the simulation tool CST Microwave Studio that was instrumental in this analysis.